\documentclass[prb,amssymb,twocolumn,superscriptaddress]{revtex4}
\usepackage{bm}
\usepackage{graphicx}
\usepackage{color}
\usepackage{amsmath}

\newcommand{\w}{\omega}
\newcommand{\wmin}{\omega_{\mathrm{min}}}
\newcommand{\ad}{a^{\dagger}}
\newcommand{\apd}{a^{\phantom{\dagger}}}
\newcommand{\bd}{b^{\dagger}}
\newcommand{\bpd}{b^{\phantom{\dagger}}}
\newcommand{\normal}{{\,}^{_\star}_{^\star}} 

\begin{document}

\title{Numerical Renormalization Group at marginal spectral density:\\
application to quantum tunneling in Luttinger liquids}

\author{Axel Freyn}
\author{Serge Florens}
\affiliation{Institut N\'eel, CNRS and Universit\'e Joseph Fourier, 
25 avenue des Martyrs, BP 166, 38042 Grenoble, France}

\begin{abstract} 
Many quantum mechanical problems (such as dissipative phase fluctuations in 
metallic and superconducting nanocircuits, or impurity scattering in Luttinger liquids) 
involve a continuum of bosonic modes with a marginal spectral density diverging 
as the inverse of energy.
We construct a Numerical Renormalization Group in this singular case, with a
manageable violation of scale separation at high energy, capturing reliably the 
low energy physics. 
The method is demonstrated by a non-perturbative solution over several energy decades 
for the dynamical conductance of a Luttinger liquid with a single static defect.
\\
\end{abstract}

\maketitle

Bosonic description of fermionic systems, possibly subject to strong
interactions, has a long history, ranging from phase fluctuations 
in superconducting circuits~\cite{Likharev,Werner} following Josephson's initial ideas,
to quantum transport in metallic grains~\cite{Schoen} and in strongly correlated
materials near the Mott transition,~\cite{Florens} where the phase conjugate to the 
electron charge is the relevant physical variable to understand the interplay of 
tunneling and Coulomb blockade. Another interesting example concerns one-dimensional 
electronic wires, the so-called Luttinger liquids (LL), where non-interacting plasmon 
modes provide a faithful representation of electronic density 
fluctuations.~\cite{Giam,Gogolin,Delft}
Quite remarkably, all these different physical problems share very common
features, because one can describe the electrons by a bosonic variable $\Phi$ conjugate to the
electronic charge transfered in a nanoscale junction (or also to the charge density
in a wire), via a phase factor $e^{i\Phi}$. This mapping can be used for
instance to represent the Josephson current in superconducting junctions (or also the 
fermionic fields in the bosonization language).
This implies in turn that the nature of the phase dynamics determines the
underlying physics:
wild fluctuations of $\Phi$ occur for instance in the presence of strong Coulomb 
blockade, leading to a rapid decay of the phase, and implying electronic 
localization.~\cite{Schoen,Florens} In contrast, phase localization is
associated to small fluctuations of the electron charge, characterizing dissipationless 
supercurrent~\cite{Likharev} or Fermi liquid states.~\cite{Florens} The intermediate 
situation of soft (algebraic) phase decay leads to the well-known non-Fermi liquid 
features of a LL.~\cite{Giam,Gogolin,Delft}

In all these situations, great complexity arises due to the coupling of the 
bosonic mode to static disorder or dynamical defects, such as 
discrete Andreev levels \cite{Zazunov,Shytov} in superconducting weak
links~\cite{Pillet}) or magnetic Kondo impurities in metallic
junctions~\cite{Hewson,Florens2} and interacting unidimensional 
wires.~\cite{Fabrizio,Egger}
Focusing the discussion on the case of impurity effects in LL, but keeping this
more general framework in mind, many technical and physical questions are still
open to date, both in the original fermionic formulation and in the bosonic version of
the problem. 
On the fermionic side, one needs to handle strong interactions within unidimensional 
wires together with the presence of exponentially small energy scales arising from the 
impurity,~\cite{Hewson,KaneFisher} for which powerful numerical methods have been developed in 
the past. The Density Matrix Renormalization Group~\cite{DMRG} can tackle
correlated wires, but on a linear energy scale only, which does not allow to really extract
critical exponents;~\cite{Meden} the Numerical Renormalization Group
(NRG)~\cite{Wilson,Bulla} can however deal with impurity physics on an exponential energy 
range, but only for uncorrelated Fermi liquids. A method that could incorporate both virtues 
would therefore be quite useful, which is the goal of this Letter.

Using the bosonic language, the description of interacting electrons by non-interacting 
bosons helps tremendously, but difficulties still arise.
Apart from perturbative analysis (or fine tuning of the model parameters to allow an exact
solution),~\cite{Giam,Gogolin,Delft} the analytical bosonization 
technique offers limited information on quantum impurity problems, because the physics crosses 
over from weak to strong coupling, for instance due to Kondo screening. Actually,
all this complexity is already encoded by a static defect in LL, which drives
the conductance from $e^2/h$ to zero on an energy scale that can be exponentially small 
in the backscattering amplitude, a problem that has triggered substantial
work, based on approximate analytical 
methods~\cite{KaneFisher,Guinea,Glazman,Enss,Meden,Aristov} or numerical techniques on 
a linear energy scale, such as quantum Monte Carlo.~\cite{Fisher,Werner}
 
The idea we henceforth present here is to use recent developments of the bosonic 
NRG~\cite{Tong1,Ingersent,Freyn} in order to tackle numerically the phase 
fluctuation problem in a broad range of parameters, with possible extensions to dynamical
defects. This however faces an immediate and seemingly intractable difficulty.
Quite common to the Josephson effect in a dissipative environment,~\cite{Shytov} to quantum tunneling 
in resistive circuits,~\cite{Safi,Florens2,Mora,Martin} or to tunneling into
LL,~\cite{KaneFisher,Guinea,Glazman,Enss,Meden,Aristov,Florens2} is the
marginal form of the bare local bosonic spectrum, given by the correlation function
$\mathcal{G}_\Phi^0(i\w) = \frac{2\pi}{|\w|}$ at imaginary frequency.
The key step in the NRG procedure is the scale separation that results from a logarithmic 
discretization of the energy band $\w_n = \w_c \Lambda^{-n}$, with $1<\Lambda$. 
A generalized power-law density of states (with exponent $s\geq-1$ and high energy cutoff $\w_c$)
of the form $J(\w)= 2\pi \w_c^{-1-s} \w^s \Theta(\w_c-\w)$ can be considered both for 
fermionic~\cite{Buxton} and bosonic models,~\cite{Tong1,Ingersent,Freyn}
providing the following coupling strength of the states at energy $\w_n$:
\begin{equation}
\gamma_n^2=\int_{\w_{n+1}}^{\w_n}\!\!\!\! d\w\, J(\w) =
2\pi \frac{1-\Lambda^{-(s+1)}}{s+1} \Lambda^{-n(s+1)} .
\label{gamma}
\end{equation}
For all $s>-1$, the couplings $\gamma_n^2$ decay exponentially with $n$,
which allows an iterative diagonalization of the problem:
the possibility of building progressively the Hilbert space from high to low energies 
is the reason behind the huge success of NRG to solve quantum impurity
problems in a linear numerical effort.~\cite{Wilson,Bulla} 
However, the marginal case $s=-1$ is special in the sense that the couplings
$\gamma_n^2= 2 \pi \log(\Lambda)$ do not decay anymore, invalidating clearly the 
whole scheme. 
We stress that we are considering quantum impurity Hamiltonians
that depend explicitly on the phase factor $e^{i\Phi}$, and not on the spatial
derivative of the bosonic mode, $\partial_x \Phi$. This latter case, which
arises for instance in the so-called ohmic spin-boson model,~\cite{Tong1} corresponds
to the much simpler situation of linear spectrum ($s=1$) of the field 
$\partial_x \Phi$, and can be easily handled by the bosonic NRG.

This complete violation of scale separation for the marginal case $s=-1$ seems
to disqualify our proposed extension of the NRG.
However, a free electron wire with constant density of states (described by the standard 
fermionic NRG at $s=0$) is equivalent to a free bosonic bath with $s=-1$ due to the 
bosonization mapping, so that one may believe that the marginal situation could
be tackled using some clever variant of the bosonic NRG. 
In order to move forward, let us investigate with greater detail the problem of
tunneling in LL. The fermionic Hamiltonian reads in terms of second quantized 
left and right moving electron modes $\psi^\dagger_{L,R}(x)$ at linear position 
$x$ in the wire (omitting the electron spin for simplicity):
\begin{eqnarray}
\nonumber
H &=& \int d x [ i v_F \psi^\dagger_L \partial_x \psi^{\phantom{\dagger}}_L 
-i v_F \psi^\dagger_R \partial_x \psi^{\phantom{\dagger}}_R + g_2
\psi^\dagger_R \psi^{\phantom{\dagger}}_R \psi^\dagger_L
\psi^{\phantom{\dagger}}_L] \\
& & - V_{bs} [ \psi^\dagger_R \psi^{\phantom{\dagger}}_L
+ \psi^\dagger_L \psi^{\phantom{\dagger}}_R]_{|x=0}
\label{Hinit}
\end{eqnarray}
where $v_F$ is the Fermi velocity, $g_2$ the short-range Coulomb repulsion between 
left and right moving electrons, and $V_{bs}$ the impurity backward scattering
amplitude at the $x=0$ location of the defect 
(forward impurity scattering and $g_4$ interaction within a given Fermi point do 
not affect the physics, and were discarded.).
The presence of the interaction term $g_2$ clearly prevents a direct fermionic NRG 
solution of the model, which requires Fermi liquid leads. Yet, one can use the
exact bosonization mapping~\cite{Giam,Gogolin,Delft} to re-express the electronic variables 
in terms of non-interacting collective charge density excitations $\Phi(x)$ and
conjugate field $\Pi(x)$. After standard
manipulations~\cite{Giam,Gogolin,Delft,KaneFisher} 
one obtains
\begin{eqnarray}
\nonumber
H\!\! &=& \!\!\! \int \frac{dx}{8\pi} \left\{ [\Pi(x)]^2 + [\partial_x \Phi(x)]^2\right\} 
- v \normal\! \cos[\sqrt{K} \Phi(x=0)]\! \normal\\
\label{Hboso}
\end{eqnarray}
where normal ordering of the cosine operator, which will be crucial for the 
rigorous formulation of the NRG algorithm, has been emphasized. We have also
introduced a small backscattering energy scale $v\propto V_{bs}$ and the important
Luttinger liquid parameter $K=[(1-g_2)/(1+g_2)]^{1/2}\leq 1$, into which all
interaction effects have been encapsulated.

Let us now present how the bosonic NRG \cite{Tong1} can be tailored to address
the impurity model~(\ref{Hboso}), which has the form of a boundary Sine Gordon
Hamiltonian. The derivation of the ``star''-NRG follows the usual
procedure~\cite{Bulla,Tong1} by considering the equivalent energy representation
in terms of a continuum of canonical bosons $\ad_\epsilon$:
\begin{eqnarray}
\label{Henergy}
H\!\! &=& \!\!\! \int_0^{\w_c} \!\!\! d\epsilon \; \epsilon \; \ad_\epsilon
\apd_\epsilon 
- v \normal\! \cos[\sqrt{K}\Phi] \!\normal,\\
\Phi & \equiv & \Phi(x=0) =  \sqrt{2} \int_0^{\w_c} \!\!\! d\epsilon \;  
\; \frac{\ad_\epsilon+\apd_\epsilon}{\sqrt{\epsilon}}.
\label{Phi}
\end{eqnarray}
The bosonic fields are then decomposed in Fourier modes ($p\in\mathbb{Z}$, $n\in\mathbb{N}$) 
on each interval $\w_{n+1}<\epsilon<\w_n$ of width
$d_n=(1-\Lambda^{-1})\Lambda^{-n}$:
\begin{eqnarray}
\label{field}
\ad_\epsilon = \sum_{n,p} \frac{e^{i 2\pi p \epsilon/d_n}}{\sqrt{d_n}}
\, a^\dagger_{n,p}.
\end{eqnarray}
The first NRG approximation consists in neglecting all $p\neq0$ modes, keeping
only the operators $\ad_n \equiv a^\dagger_{n,0}$ (this step becomes exact in the
$\Lambda\to1$ limit~\cite{Bulla}). This leads to the ``star''-Hamiltonian:
\begin{eqnarray}
\label{Hstar}
\!\!\!\! H_S\!\! &=& \!\!\! \sum_{n=0}^{+\infty} \xi_n \ad_n \apd_n
- v \normal \! \cos\left[\sqrt{K} \sum_{n=0}^{+\infty} 
\frac{\gamma_n}{\sqrt{\pi}} (\ad_n+\apd_n)\right]\!\!\!\normal
\end{eqnarray}
with the ``impurity'' coupling strength already given in Eq.~(\ref{gamma}) by 
$\gamma_n^2=2\pi\log(\Lambda)$ in the marginal case $s=-1$. 
The typical energy $\xi_n$ in each shell is defined by:
\begin{equation}
\xi_n = \frac{1}{\gamma_n^2} \int_{\w_{n+1}}^{\w_n}\!\!\!\! d\w\, \w\, J(\w) =
\frac{1-\Lambda^{-1}}{\log(\Lambda)} \w_c \Lambda^{-n}.
\label{xi}
\end{equation}
As a benchmark of the discretization for the marginal case $s=-1$, one can
easily compute from~(\ref{Hstar}) the resulting approximation for the original 
Green's function:
\begin{equation}
\mathcal{G}_{\Phi,\Lambda}^{0}(i\w) = \frac{4}{1-\Lambda^{-1}} \sum_{n=0}^{+\infty}
\frac{\w_c \Lambda^{-n}}{\w^2+\left[\frac{1-\Lambda^{-1}}{\log(\Lambda)}\right]^2
\w_c^2 \Lambda^{-2n}}
\end{equation}
which can be checked to converge exponentially fast at $\w\ll\w_c$ to the exact result
$\mathcal{G}_\Phi^0(i\w)=\frac{2\pi}{|\w|}$ even for $\Lambda=2$ (we keep this standard
value from now on).
However, despite the clear exponential decay of the energies~(\ref{xi}),
the non-decreasing value of the couplings $\gamma_n$ implies a violation of scale
separation on {\it all} shells, and prevents the solution by iterative diagonalization
of Hamiltonian~(\ref{Hstar}).

\begin{figure}[ht]
\includegraphics[width=8.7cm]{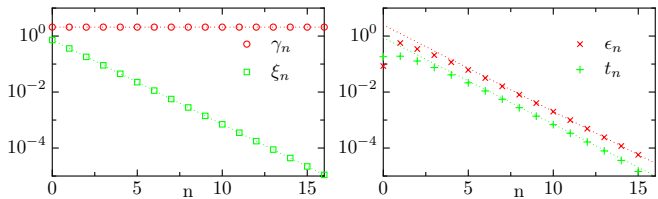}
\caption{(color online) Left panel: parameters $\xi_n$ and $\gamma_n$ of the
star-NRG as a function of $n$ for $0\leq n \leq n_\mathrm{min}=16$; the coupling $\gamma_n$ does
not decay and violates scale separation on {\it all} shells. Right panel: 
parameters $\epsilon_n$ and $t_n$ of the chain-NRG; scale separation is only 
broken on the first shell, as seen by the initial increase of both parameters, 
before further exponential decay (shown by dotted lines as guides to the eye).}
\label{params}
\end{figure}

The first key idea in successfully constructing the marginal bosonic NRG is to assume 
that the energy spectrum is also bounded from {\it below}:
\begin{equation}
J(\w) = \frac{2\pi}{\w} \Theta(\w_c-\w) \Theta(\w-\wmin) .
\end{equation}
Clearly both the energies $\epsilon_n$ and the couplings $\gamma_n$ are not
modified by this choice ($\gamma_n$ still do not decay), and they are just
cut off for $n>n_\mathrm{min}$, with $\wmin=\w_c\Lambda^{-n_\mathrm{min}}$, 
so that nothing seems gained naively. We can however try to pursue with the second step of the 
standard NRG procedure, which amounts to the exact mapping on the Wilson
chain.~\cite{Bulla,Tong1}
This simple tridiagonalization procedure of Hamiltonian~(\ref{Hstar}) leads to the 
following ``chain'' form in terms of new canonical bosons $\bd_n$:
\begin{eqnarray}
\nonumber
H_C\!\! &=& \!\!\! \sum_{n=0}^{+\infty} \left[ \epsilon_n \bd_n \bpd_n
+ t_n (\bd_n \bpd_{n+1}+\bd_{n+1} \bd_n) \right]\\
&& - v \normal \! \cos\left[ \sqrt{\frac{\eta_0 K}{\pi}} (\bd_0+\bpd_0)
\right] \!\!\! \normal
\label{Hchain}
\end{eqnarray}
with the parameter $\eta_0=2\pi\log(\w_c/\wmin)$. 
Clearly, the impurity part of the chain Hamiltonian~(\ref{Hchain}) breaks down 
for $\wmin\to0$, owing to the divergence of $\eta_0$, but one can check
numerically that the construction is valid for non-zero $\wmin$. 
The on-site energies $\epsilon_n$ and hoppings $t_n$ of the Wilson chain
can indeed be obtained by numerical tridiagonalization of Eq.~(\ref{Hstar}).
For the value $\wmin=10^{-5}$ of the lower cutoff, these are plotted together 
with the star parameters on Fig.~\ref{params}.
The exponential decay of {\it both} chain parameters $\epsilon_n$ and $t_n$ that we 
discover here is clearly a remarkable surprise, that enables the extension of the NRG 
to the marginal situation $s=-1$. This crucial feature comes at a 
small price, seen by the first increase of the chain parameters from site $n=0$ to site 
$n=1$. Thus the maximal violation of scale separation in the star NRG presents
a small remanence in the chain NRG, limited only to the first shell. Interestingly, 
the initial jump of the parameters is just proportional to $\log(\w_c/\wmin)$, 
so that the lower cutoff $\wmin$ can be decreased on exponential scales without paying
a huge numerical cost.

A last difficulty due to the unusual form of the impurity Hamiltonian~(\ref{Hchain}) 
must be addressed. In the standard
NRG,~\cite{Bulla,Tong1} only linear to quadrilinear operators are present
in the Hamiltonian. However, the central physical role played by the phase
factor $e^{i\Phi}$ leads to a cosine term at the impurity site, hence to an operator of 
infinite order, which by the bosonization rules~\cite{Delft} must also be normal 
ordered. For a generic operator $\mathcal{O} = \normal \cos[\alpha(\bd_0+\bpd_0)]\normal$
this reads $\mathcal{O} = \cos(\alpha\bd_0) \cos(\alpha \bpd_0) - 
\sin(\alpha\bd_0) \sin(\alpha \bpd_0)$. Using the Fock states $\big|m\big>$ of the 
bosonic creation operator $\bd_0$ on the initial site $n=0$ of the Wilson chain,
one obtains the matrix elements:
\begin{equation}
\big <m\big|\mathcal{O}\big|p\big> = \sqrt{m! p!} \, \mathcal{R}e
\sum_{k=0}^{\mathrm{Min}(m,p)}  \!\!\!\!
\frac{(i\alpha)^{m+p-2k}}{(m-k)! (p-k)! k!} .
\label{matrix}
\end{equation}
The construction of the impurity term in~(\ref{Hchain}) proceeds by a truncation
of the infinite Fock space on the initial Wilson site limited to states with occupation
number less than a given $N_0$, and use of the matrix elements~(\ref{matrix}). Typically
$N_0=150$ ensures a good representation of the Hamiltonian.
Each further site $n>0$ of the chain is described by a basis of $N_b$ states~\cite{Tong1} 
(we take $N_b=12$ here). At increasing $n$, the growing size of the total 
Hilbert space becomes rapidly unmanageable, and a truncation 
to $N_\mathrm{trunc}$ states is required (this approximation is common to all 
NRG schemes~\cite{Bulla}). Typically $N_\mathrm{trunc}=800$ was 
employed in all further computations, and we also set $\w_c=1$ as the basic
energy unit.

\begin{figure}[ht]
\includegraphics[width=8.0cm]{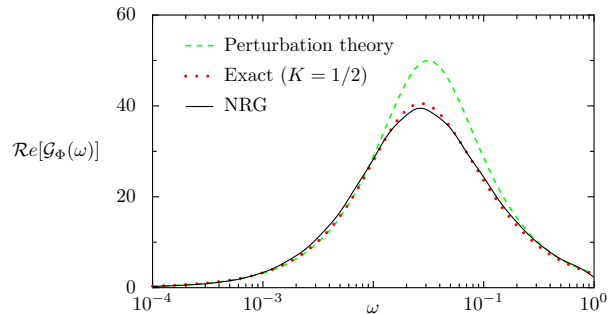}
\caption{(color online). Bosonic correlation function
$\mathrm{R}e[\mathcal{G}_\Phi(\w)]$ at real frequency $\w$ for the LL parameter
$K=1/2$ comparing (bottom to top) the NRG to the exact result~(\ref{solvable}),
and to the strong and weak interaction perturbation theory given 
respectively by~(\ref{strong}) and~(\ref{weak}) (these two expressions are 
by accident equivalent for $K=1/2$, but nonetheless not exact).}
\label{TestGPhi}
\end{figure}
\begin{figure}[ht]
\includegraphics[width=8.0cm]{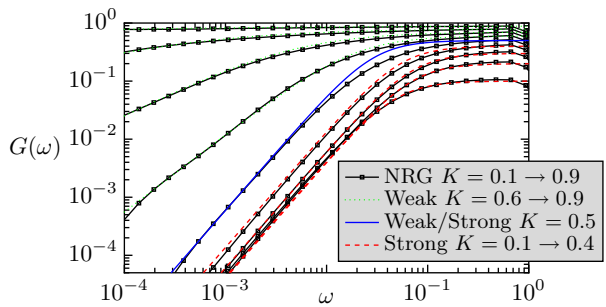}
\caption{(color online).
Dynamical conductance $G(\w)$ in units of $e^2/h$ for several values of the LL 
parameter $K=0.1,0.2,0.3,0.4,0.5,0.6,0.7,0.8,0.9$ obtained by the 
NRG (symbols, bottom to top). Comparison is made for $K\leq0.5$ to the
strong interaction and for $K\geq0.5$ to the weak interaction
limits. Perturbative methods cannot describe accurately the NRG solution beyond
their applicability range.}
\label{Conductance}
\end{figure}
In contrast to more complex extensions of impurity models with dynamical 
degrees of freedom (such as the Kondo model in a Luttinger
liquid~\cite{Fabrizio,Egger}), the present impurity problem benefits from several
known limits, that allow to benchmark our numerics.
For instance there exists an exact solution for the dynamical
conductance~\cite{KaneFisher,Giam,Gogolin,Delft} (in units of $e^2/h$) at $K=1/2$:
\begin{equation}
\label{solvable}
 G^\mathrm{exact}(\w) = \frac{K \w}{2\pi} \mathcal{R}e [\mathcal{G}_\Phi(\w)] =
\frac{1}{2}-\frac{\Omega}{2\w}\mathrm{atan}\left(\frac{\w}{\Omega}\right)
\end{equation}
where $\Omega=e^\gamma v^2$, with Euler's constant $\gamma$, is the crossover energy 
at which the impurity cuts the chain (for $K=1/2$). Perturbation theory works
also at strong interaction $K\ll1$, in which case the self-consistent
harmonic approximation applies:~\cite{Guinea}
\begin{equation}
\label{strong}
G^\mathrm{strong}(\w) = \frac{1}{2}\frac{\w^2}{\w^2+(\Omega^\star)^2}
\end{equation}
with $\Omega^\star=2\pi K v^{\frac{1}{1-K}}$ the crossover scale.
Finally, the limit of weak interaction $1-K\ll1$ is also 
known from several approaches:~\cite{KaneFisher,Glazman,Enss,Meden,Giam,Gogolin,Aristov}
\begin{equation}
\label{weak}
G^\mathrm{weak}(\w) =
\frac{K \w^{\frac{2}{K}-2}}{\w^{\frac{2}{K}-2}+(\Omega^\star)^{\frac{2}{K}-2}}.
\end{equation}
Fig.~\ref{TestGPhi} compares our NRG data for $K=1/2$ with the exact solution and
perturbation theory, which shows the excellent convergence of the NRG and the sizeable 
discrepancies of both perturbative expansions. More systematic analysis for
various $K$ values in Fig.~\ref{Conductance} demonstrates the progressive departure
of the perturbative results from the numerical solution. The ability of the
marginal bosonic NRG to describe non-perturbatively universal transport features 
with {\it high accuracy} should therefore make it a precious tool to test scaling 
behavior of impurity physics in LL.

To conclude, we have established an extension of the NRG to deal with the marginal 
situation of a density of states diverging as the inverse of energy.
The potentially most promising applications of the NRG at marginal coupling
concern the physics of dynamical impurities coupled to phase fluctuations, a
large class of physical problems where no alternative analytical or numerical techniques
exist to date. This development could allow to address many currently open issues, such 
as non-equilibrium transport with strong correlations (using a mapping onto equilibrium 
q-oscillator models~\cite{Lukyanov}), Kondo physics in Luttinger
liquids,~\cite{Fabrizio,Egger,Florens2} and ohmic dissipation in Andreev level 
qubits.~\cite{Zazunov,Shytov,Pillet}

We thank L. Glazman and V. Meden for useful discussions. Financial support 
from ERC Advanced Grant MolNanoSpin No. 226558 is also gratefuly acknowledged.


\begin{thebibliography}{50}

\bibitem{Likharev} K. K. Likharev, {\it Dynamics of Josephson Junctions and
Circuits} (Gordon, 1986).
\bibitem{Werner} S. L. Lukyanov and P. Werner, J. Stat. Mech. {\bf 06}, 06002
(2007).

\bibitem{Schoen} G. Sch\"on and A. D. Zaikin, Phys. Rep. {\bf 198}, 237 (1990).

\bibitem{Florens} S. Florens and A. Georges, Phys. Rev. B {\bf 70}, 035114 (2004).

\bibitem{Giam} T. Giamarchi, {\it Quantum physics in one dimension} (Oxford,
2003).
\bibitem{Gogolin} A. O. Gogolin, A. A. Nersesyan, and A. M. Tsvelik, 
{\it Bosonization and strongly correlated systems} (Cambridge, 1998).
\bibitem{Delft} J. von Delft and H. Schoeller, Ann. Phys. {\bf 7}, 225 (1998).

\bibitem{Zazunov} A. Zazunov, V. S. Shumeiko, G. Wendin, and E. N. Bratus, Phys.
Rev. B {\bf 71}, 214505 (2005).
\bibitem{Shytov} A. V. Shytov, arXiv:cond-mat/0001012.
\bibitem{Pillet} J.-D. Pillet {\it et al.}, Nature {\bf 6}, 965
(2010).

\bibitem{Hewson} A. C. Hewson, {\it The Kondo Problem to Heavy
Fermions} (Cambridge, 1996).
\bibitem{Florens2} S. Florens, P. Simon, S. Andergassen, and D. Feinberg,
Phys. Rev. B {\bf 75}, 155321 (2007).

\bibitem{Fabrizio} M. Fabrizio, and A. O. Gogolin, Phys. Rev. B {\bf 51}, 17827
(1995).
\bibitem{Egger} R. Egger and A. Komnik, Phys. Rev. B {\bf 57}, 10620 (1998).

\bibitem{KaneFisher} C. L. Kane, and M. P. A. Fisher, Phys. Rev. Lett. {\bf
68}, 1220 (1992); Phys. Rev. B {\bf 46}, 7268 (1992).

\bibitem{DMRG} S. R. White,  Phys. Rev. Lett. {\bf 69}, 2863 (1992).
\bibitem{Meden} V. Meden {\it et al.}, 
New J. Phys. {\bf 10}, 045012 (2008).

\bibitem{Wilson} H.R. Krishna-murthy, J.W. Wilkins, and K.G. Wilson Phys. Rev.
B {\bf 21}, 1003 (1980).
\bibitem{Bulla} R. Bulla, T. Costi and T. Pruschke, Rev. Mod. Phys. {\bf 80},
395 (2008).

\bibitem{Guinea} F. Guinea, G. Gomez Santos, M. Sassetti, and M. Ueda, Euro.
Phys. Lett. {\bf 30}, 561 (1995).
\bibitem{Glazman} D. Yue, L. I. Glazman, and K. A. Matveev, Phys. Rev. B 
{\bf 49}, 1966 (1994).
\bibitem{Enss} T. Enss {\it et. al}, Phys. Rev. B {\bf 71}, 155401 (2005).
\bibitem{Aristov} D. N. Aristov, and P. Woelfle, Euro. Phys. Lett. {\bf 82},
27001 (2008).
\bibitem{Fisher} K. Moon, H. Yi, C. L. Kane, S. M. Girvin, and M. P. A. Fisher,
Phys. Rev. Lett. {\bf 71}, 4381 (1993).

\bibitem{Tong1} R. Bulla, N.-H. Tong, and M. Vojta, Phys. Rev. Lett. {\bf 91}, 
170601 (2003);  R. Bulla, H.-J. Lee, N.-H. Tong and M. Vojta, Phys. Rev. B 
{\bf 71}, 045122 (2005).
\bibitem{Ingersent} M. T. Glossop, and K. Ingersent, Phys. Rev. Lett. {\bf 95},
067202 (2005).
\bibitem{Freyn} A. Freyn and S. Florens, Phys. Rev. B {\bf 79}, 121102 (2009).

\bibitem{Safi} I. Safi and H. Saleur, Phys. Rev. Lett. {\bf 93}, 126602 (2004).
\bibitem{Mora} C. Mora, and K. Le Hur, Nature Phys. {\bf 6}, 697 (2010).
\bibitem{Martin} Y. Hamamoto, T. Jonckheere, T. Kato, and T. Martin,
Phys. Rev. B {\bf 81}, (2010) 153305.

\bibitem{Buxton} C. Gonzalez-Buxton, and K. Ingersent, Phys. Rev. B {\bf 57}, 
14254 (1998).

\bibitem{Lukyanov} V. Bazhanov, S. Lukyanov, and A. Zamolodchikov,
Nucl. Phys. B {\bf 549}, 529 (1999).

\end{thebibliography}
\end{document}